\documentclass[11pt]{article}
\usepackage{amsmath}
\usepackage{amssymb}
\usepackage{amsthm}

\def\fpd#1#2{{\displaystyle\frac{\partial #1}{\partial #2}}}

\def\lie#1{{\mathcal{L}}_{#1}}

\def\del0Z{{\nabla}_{\!Z_0}}
\def\deltZ{{\nabla}_{\!\tilde{Z}_0}}
\def\delZ{{\nabla}\!Z_0}

\def\R{\mathbb{R}}
\def\Im#1{{\mbox{Im}}\,{#1}}
\def\barE{\overline{E}}
\def\sp{\mathop{\rm sp}}
\def\rot{\mathop{\rm rot}}
\def\divg{\mathop{\rm div}}

\def\onehalf{{\textstyle\frac12}}
\def\onequarter{{\textstyle\frac14}}

\def\hook{\vrule height 0pt depth 0.4pt width 3pt \vrule height 5pt depth 0.4pt
\kern 3pt}

\newtheorem{thm}{Theorem}
\newtheorem{lem}{Lemma}
\newtheorem{prop}{Proposition}

\begin{document}

\title{A generalization of Szebehely's inverse problem of dynamics in dimension three}
\author{W.\ Sarlet$^{a,b}$, T.\ Mestdag$^{a,c}$ and G.\ Prince$^b$ \\[2mm]
{\small ${}^a$Department of Mathematics, Ghent University }\\
{\small Krijgslaan 281, B-9000 Ghent, Belgium}\\[1mm]
{\small ${}^b$Department of Mathematics and Statistics, La Trobe University}\\
{\small Melbourne, Victoria 3086, Australia}\\[1mm]
{\small ${}^c$Department of Mathematics and Computer Science, University of Antwerp}\\
{\small Middelheimlaan 1, B-2020 Antwerp, Belgium}\\[1mm]
{\small (e-mails: willy.sarlet@ugent.be,\ \ mestdagtom@gmail.com,\ \ geoff@amsi.org.au)}
}

\date{}

\maketitle

\begin{quote}
{\small {\bf Abstract.} Extending a previous paper, we present a generalization in dimension 3 of the traditional Szebehely-type inverse problem. In that traditional setting, the data are curves determined as the intersection of two families of surfaces, and the problem is to find a potential $V$ such that the Lagrangian $L = T - V$, where $T$ is the standard Euclidean kinetic energy function, generates integral curves which include the given family of curves. Our more general way of posing the problem makes use of ideas of the inverse problem of the calculus of variations and essentially consists of allowing more general kinetic energy functions, with a metric which is still constant, but need not be the standard Euclidean one. In developing our generalization, we review and clarify different aspects of the existing literature on the problem and illustrate the relevance of the newly introduced additional freedom with many examples.}
\end{quote}

{\bf Keywords}: Szebehely's equation, inverse problem of dynamics,
inverse problem of the calculus of variations

\section{Introduction}

In a previous paper \cite{SMP}, we observed that the extensive literature on the so-called Szebehely inverse problem of dynamics (sometimes also referred to as Suslov's or Joukovsky's problem, see e.g.\ \cite{Shorokov}) apparently has overlooked a more general way of posing the problem, leading to more general solutions as well as cases that have no solutions otherwise. Our analysis made clear that the nature of the problem very much depends on the dimension $n$ of the underlying space and we restricted ourselves in \cite{SMP} for this reason to the case $n=2$. We refer to \cite{SMP} for general references and comments about the history of the problem. In the present paper, about the case $n=3$, references will be restricted as much as possible to those which are of direct relevance to our analysis or examples.

For completeness, we start with a brief sketch of the traditional problem in $\R^3$ (see e.g.\ \cite{Bozis1995}) and again explain the motivation for our proposed generalization. We implicitly assume that all functions are smooth on some open set of $\R^3$. Consider a family of curves determined by two given families of surfaces, specified as level sets, say
\begin{equation}
\phi(x,y,z) = c_1, \qquad \psi(x,y,z)=c_2, \quad \mbox{with}\quad d\phi\wedge d\psi\neq 0. \label{data}
\end{equation}
The traditional question is to find a potential function $V(x,y,z)$ such that the set of integral curves of the classical mechanical system
\begin{equation}
\ddot x = - \fpd{V}{x}, \quad \ddot y = - \fpd{V}{y}, \quad \ddot z = - \fpd{V}{z}, \label{oldSz}
\end{equation}
contains the given family (\ref{data}). Equivalently, the idea is to construct a Lagrangian $L=T - V$ whose Euler-Lagrange equations have the given family as part of its integral curves. But why should the kinetic energy function in such a construction come from the standard Euclidean metric? Our suggestion therefore was to bring this other well-known inverse problem into the picture, namely the question (with index notation now): for a second-order dynamical system $\ddot{x}^i = F^i(x,\dot{x})$ with given forces $F^i$, find a non-singular symmetric multiplier matrix $g_{ij}$ such that
\begin{equation}
g_{ij} (\ddot x^j - F^j) \equiv \frac{d}{dt} \left(\frac{\partial
L}{\partial \dot x^i}\right) - \frac{\partial L}{\partial x^i}, \label{multiplier}
\end{equation}
for some Lagrangian function $L(x,\dot x)$. In view of the data in the current context, we consider only the case that the forces do not depend on the velocities and equally restrict the unknown multiplier matrix to one which does not depend on velocities. As explained in \cite{SMP}, it then follows that the $g_{ij}$ actually have to be constants. Note that there have been studies in the literature where the Szebehely type problem was addressed with a non-Euclidean and non-constant metric $g_{ij}(x)$ in the kinetic energy term (see e.g.\ \cite{Melis-Piras1984}). But in those studies, the metric was assumed to be given from the outset and of course, through the connection coefficients of the Levi-Civita connection, it gives rise to extra force terms which are quadratic in the velocities. So there is room for further generalizations, in principle, but once the idea is that given forces are not strictly dictated by an as yet undetermined potential, this becomes an entirely different story all together. The interest of the generalization we propose is that it is still about purely conservative forces, but introduces (for $n=3$) six additional free parameters which can significantly enlarge the class of solvable problems: essentially, from the point of view of finding a potential $V$, the requirement
\[
F^i(x) = - \fpd{V}{x^i} \quad \mbox{is replaced by} \quad g_{ij}F^j(x) = - \fpd{V}{x^i},
\]
for some symmetric, non-degenerate matrix $g_{ij}$.

The scheme of the paper is as follows. In Section~2, we present details of the formulation of the problem and our general approach to it in the generic case that a constant potential is not a solution, i.e.\ that the given curves (\ref{data}) are not all straight lines. We also discuss one of the possible ways of addressing the problem, which amounts to solving one first- and one second-order partial differential equation for $V$. In Section~3, a full integrability analysis of the original system of first-order equations is carried out, leading to a classification in various subcases. We further explain how the presence of the six additional unknown parameters in our formulation makes this integrability scheme particularly relevant. In Section~4, we prove that one of the naturally arising subcases in our integrability classification turns out to be void. A representative selection of examples from the literature is discussed from our broader perspective in Section~5. We present some new examples in Section~6, with particular attention for cases which are shown to have a solution in our generalized approach, while having no solution in the standard Euclidean case.

\section{The amalgamated inverse problem for $n=3$}

A preliminary remark concerning our approach: although the problem to be addressed in the end boils down to solving partial differential equations, we think it is important to work with well defined geometrical objects in the formulation of the problem, so that its nature does not change when we pass to other coordinate representations.

A vector field $Z$ on $\R^3$ is tangent to the curves (\ref{data}) if it satisfies $Z(\phi)=Z(\psi)=0$. The way to ensure that the given curves become (projections of) integral curves of an associated second-order vector field $\Gamma$ on the tangent bundle $T\R^3$ goes as follows. The lifts to $T\R^3$ of the given curves are integral curves of the complete lift $Z^c$ of $Z$, which has the coordinate expression
\[
Z^c = Z^i(x)\fpd{}{x^i} + \dot{x}^j\fpd{Z^i}{x^j}\fpd{}{\dot{x}^i}.
\]
However, $Z^c$ is not a second-order vector field; such a vector field is of the form
\[
\Gamma = \dot{x}^i\fpd{}{x^i} + F^i\fpd{}{\dot{x}^i}
\]
for some `forces' $F^i$, and the requirement that its integral curves contain the relevant ones of $Z^c$ amounts to imposing that
\begin{equation}
Z^c|_{\Im{Z}} = \Gamma|_{\Im{Z}} \quad \mbox{or explicitly} \quad Z(Z^i)= Z^j\fpd{Z^i}{x^j}= F^i|_{\Im{Z}}. \label{ZGamma}
\end{equation}
Since the $Z(Z^i)$ are functions of the $x^i$ only we can restrict ourselves to forces $F^i$ with the same property and then we can simply write $F^i$ in (\ref{ZGamma}).

The next element in our construction is that $\Gamma$ should be the normal form representation of a set of Euler-Lagrange equations. The nature of the functions $F^i$ we have in mind entails that the as yet undetermined kinetic energy metric $g$ must be constant and together with the unknown potential must satisfy the relations
\begin{equation}
g_{ij}Z(Z^j) = - V_{x^i}. \label{Vxi1}
\end{equation}
We now pass to a basis consisting of a vector tangent to the given curves and two others spanning the orthogonal complement (with respect to the unknown $g$). The vector $Z$ we start from can be any tangent vector, so we put $Z= h Z_0$ and choose $Z_0$ to be the vector field
\begin{equation}
Z_0 = (\phi_y \psi_z - \phi_z \psi_y)\fpd{}{x} + (\phi_z \psi_x - \phi_x \psi_z)\fpd{}{y} + (\phi_x \psi_y - \phi_y \psi_x)\fpd{}{z}. \label{Z0}
\end{equation}
In more geometrical terms, if $\omega_0$ denotes the standard volume form on $\R^3$, $Z_0$ is determined by
\begin{equation}
i_{Z_0} \omega_0 = d\phi\wedge d\psi. \label{Z0bis}
\end{equation}
A few comments are in order here. Perhaps it would seem more appropriate in the present context to use the volume form $\omega = \sqrt{\det g}\,\omega_0$ for defining $Z_0$. It also may look appealing to actually normalize $Z_0$ to be a unit vector with respect to $g$, or alternatively, to make it the Euclidean unit tangent vector and continue the subsequent analysis in terms of the Frenet frame (as was done, for example, by Puel \cite{Puel1992} for the standard problem). For computational reasons, however, we avoid normalizing various vector fields because it leads to coefficients with unwieldy denominators in partial differential equations. In fact, for the same reason, we do not stick to the specific form (\ref{Z0}) in applications either, but freely make use of rescalings when appropriate.

Two vectors which span $Z_0^\perp$, the orthogonal complement to $Z_0$, are defined by
\begin{equation}
Z_1 \hook g = d\phi, \qquad Z_2 \hook g = d\psi. \label{Z1Z2}
\end{equation}
The coordinate expression of their components is given by
\[
Z_1^i = g^{ik} \phi_{x^k}, \qquad Z_2^i = g^{ik} \psi_{x^k},
\]
and we have the obvious properties
\begin{equation}
g(Z_0,Z_1) = g(Z_0,Z_2) = 0, \qquad g(Z_1,Z_2) = Z_2(\phi) = Z_1(\psi). \label{Zprops}
\end{equation}
Since the metric $g$ is constant, the functions $Z_0(Z_0^i)$ appearing in (\ref{Vxi1}) are simply the components of the covariant derivative $\nabla_{\!Z_0}Z_0$, which we will write as $\delZ$ if there is no danger for ambiguity.

An equivalent representation of the conditions (\ref{Vxi1}), which amounts to passing from the basis of coordinate vector fields to the basis $\{Z_0,Z_1,Z_2\}$, is obtained by multiplying the equation for $V_{x^i}$ with the $i$-th component of each of these vector fields and summing over~$i$. Knowing that $Z(Z^j) = \onehalf Z_0(h^2) Z_0^j + h^2 Z_0(Z_0^j)$, the combination leading to $Z_0(V)$ gives
\begin{equation}
Z_0(\onehalf h^2 g(Z_0,Z_0) + V) = 0. \label{Z0eqn}
\end{equation}
Proceeding in the same way for $Z_1$ and $Z_2$, we can write the remaining conditions in the form
\begin{align}
Z_1(V) &= - h^2 \delZ(\phi), \label{Z1eqn} \\
Z_2(V) &= - h^2 \delZ(\psi), \label{Z2eqn}
\end{align}
where we have used the fact that
\begin{equation}
g(Z_1,\delZ) = \delZ(\phi) \quad \mbox{and}\quad g(Z_2,\delZ) = \delZ(\psi). \label{delZprops}
\end{equation}
If a solution exists, and that involves finding a $g$ and a $V$ and fixing the parametrization $h$, it is clear that the condition (\ref{Z0eqn}) expresses that the total energy function $E:=\onehalf g_{ij}\dot{x}^i\dot{x}^j + V$, when restricted to the image of the vector field $Z$, must be a function, possibly constant, of $\phi$ and $\psi$:
\begin{equation}
E|_{\Im{Z}} = \barE(\phi,\psi). \label{barE}
\end{equation}
In dimension 2, there is only one equation such as (\ref{Z1eqn}) or (\ref{Z2eqn}). The procedure to arrive at a Szebehely-type equation then consists of solving that equation for $h^2$ and substituting the result into the integrated form of (\ref{Z0eqn}). Clearly, the situation in dimension~3 is different, because there are two equations for $h^2$ and naturally these have to be compatible. Leaving aside the exceptional situation that both right-hand sides of (\ref{Z1eqn},\ref{Z2eqn}) vanish identically, which will be shown to correspond to the case of straight lines in the next section, the compatibility requirement is that
\begin{equation}
X(V) = 0, \qquad \mbox{where} \qquad X:= \delZ(\psi) Z_1 - \delZ(\phi) Z_2. \label{X}
\end{equation}
As an aside it trivially follows from (\ref{Zprops}) that $g(X,Z_0)=0$ and from (\ref{delZprops}) that $g(X,\delZ)=0$. In other words, whenever $Z_0$ and $\delZ$ are independent, they span the orthogonal complement of $X$.

The equation $X(V)=0$ is a linear, homogeneous first-order partial differential equation for V. But of course, in our generalized formulation of the problem, the vector field $X$ depends on the as yet undetermined constant multiplier matrix $(g_{ij})$. In principle, if this equation can be integrated, and rescaling $X$ may again be helpful for that purpose, one should solve the condition (\ref{Z1eqn}) for $Z_1(V)$ (or (\ref{Z2eqn}) for $Z_2(V)$) algebraically  for $h^2$. The, after substitution of that expression into the result about the energy function, try to pin down the remaining freedom in $V$ and $g$ by addressing the newly obtained linear (but inhomogeneous) first-order partial differential equation for $V$, which moreover contains the as yet undetermined function $\barE$. Defining
\begin{equation}
A:= \onehalf\, \frac{g(Z_0,Z_0)}{\delZ(\phi)}, \label{A}
\end{equation}
then this remaining equation is
\begin{equation}
A\, Z_1(V) = V - \barE. \label{AZ1}
\end{equation}

A full integrability analysis of the equations (\ref{X}) and (\ref{AZ1}) is presented in the next section. But some further comments are in order, first about the dependence on the dimension of the space. It is clear that the whole picture will be very different for $n>3$. There would then be $n-2$ linear, homogeneous equations of the type (\ref{X}),  requiring a separate, preliminary integrability analysis.

There is a direct approach to solving the problem which, if implementable, makes an integrability analysis redundant. Indeed, after solving (\ref{X}) and obtaining an expression for $h^2$, one can substitute it directly into (\ref{Z0eqn}) (instead of using the integrated form (\ref{AZ1})); this generates a second-order equation for $V$ of the form
\begin{equation}
A\,Z_0Z_1(V) = Z_0(V) - Z_0(A)\,Z_1(V). \label{Z0Z1}
\end{equation}
Following the terminology introduced in \cite{Anisiu2005} for the standard Euclidean metric, we will call this the `energy free approach'. The way to search for the solution then goes as follows: (i) solving $X(V)=0$ specifies $V$ as an arbitrary function of two variables, $(u,v)$ say; (ii) selecting a third variable $w$ in an appropriate way, a coordinate change from $(x,y,z)$ to $(u,v,w)$ may transform the second-order equation (\ref{Z0Z1}) into a polynomial expression in $w$ whose coefficients have to vanish separately and produce this way a number of conditions which lead to a solution for $V$. If this procedure works, it should of course be implemented. In practice, however, there will often be a number of computational obstacles. To begin with, in our case with a non-Euclidean but constant $g$, the presence of up to six free parameters in the coordinate expression of the vector field $X$ makes it very unlikely that the variables $(u,v)$ can be identified. And even if that works, it may be impossible to choose a $w$ in such a way that the transformation can explicitly be inverted to express the second-order condition in new variables. The integrability analysis of the next section then becomes very relevant for unravelling alternative ways of trying to obtain solutions.

\section{General integrability scheme}

As far as we know, there have been two attempts to develop an integrability classification in the existing literature for $g_{ij}=\delta_{ij}$. The first one by V\'aradi and \'Erdi \cite{VE83} is ineffective, unfortunately, if only because these authors regarded the energy function as being a priori given, an assumption which is simply inconsistent with the nature of all requirements. In their examples, a choice for what we call $\barE$ comes out of the blue and is not justified by any theoretical foundation. The integrability analysis by Shorokhov \cite{Shorokov} in this respect is much more reliable and in fact contains a surprising strong statement (cf.\ our next section) which perhaps has not been sufficiently recognised. Nevertheless, it is probably not an optimal approach either, because the two starting equations (taken over from \cite{VE83}) both contain the function $\barE$, i.e.\ there is no direct role for the particular vector field $X$ in the plane orthogonal to the given curves which has the advantage of producing a homogeneous partial differential equation not involving the energy function.

Recall first that setting up the equation for $X$  requires that the right-hand sides of (\ref{Z1eqn},\ref{Z2eqn}) do not vanish simultaneously. We will label the rather exceptional situation that they do as Case~0.

{\bf Case 0:}\ \ $\delZ(\phi) = \delZ(\psi) = 0$.

Obviously, this implies that $\delZ$ is proportional to $Z_0$ and we will show first that it corresponds to the case that the given family of curves consists of straight lines.

\begin{prop} The family of curves defined by (\ref{data}) consists of straight lines, if and only if $\,\delZ = \rho Z_0$ for some function $\rho$.
\end{prop}
{\sc Proof:}\ We know that $Z_0$ is tangent to the curves (\ref{data}). This means that along each such curve $x^i(s)$, if it is parameterized by Euclidean arclength $s$, $Z_0$ is actually proportional to $d/ds$ at each point, in other words:
\[
Z_0^i = \lambda \frac{dx^i}{ds} \quad\mbox{and}\quad Z_0(Z_0^i) = \lambda \frac{d\lambda}{ds} \frac{dx^i}{ds} + \lambda^2 \frac{d^2 x^i}{ds^2} .
\]
Hence, if $Z_0(Z_0^i) = \rho Z_0^i$, it follows that
\[
\frac{d^2 x^i}{ds^2} = \frac{1}{\lambda} \Big(\rho - \frac{d\lambda}{ds}\Big) \frac{dx^i}{ds}.
\]
But this further tells us that
\[
\frac{dx^i}{ds} \frac{dx^i}{ds} = 1 \quad \Rightarrow \quad 0 = \frac{d^2 x^i}{ds^2}\frac{dx^i}{ds} = \frac{1}{\lambda} \big(\rho - \frac{d\lambda}{ds}\big).
\]
The conclusion is that $d^2x^i/ds^2 =0$, meaning that the curvature is zero at each point and hence that we are looking at straight lines.

Conversely, if the data are such that all curves (\ref{data}) are straight lines, the above pointwise calculation shows that we will have
\[
Z_0(Z_0^i) = \lambda \frac{d\lambda}{ds} \frac{dx^i}{ds} = \frac{d\lambda}{ds} Z_0^i,
\]
and hence that $\delZ$ is proportional to $Z_0$. \qed

For straight lines, our problem obviously has the solution  $V\! = constant$, which is not terribly interesting because in such a case the extra freedom introduced by the metric is irrelevant. But the above result at least tells us that, if we are not in Case~0 and if the conditions on the potential can only be satisfied by a constant $V$, then there is actually no solution! As for Case~0, continuation of an integrability analysis reveals two subcases.

From (\ref{Z1eqn},\ref{Z2eqn}), it follows that we must have
\[
Z_1(V) = Z_2(V) = 0.
\]

{\bf Case 0a:}\ \ $Z_1$ and $Z_2$ span an integrable distribution.

Then there exists a solution for $V$ in the form of an arbitrary function of one specific variable. The remaining requirement
\[
\onehalf\,h^2 g(Z_0, Z_0) + V = \barE(\phi,\psi)
\]
can simply be viewed subsequently as determining $h^2$ as a function of $\barE(\phi,\psi)$. Hence, although the given curves are straight lines, the problem has non-constant solutions for $V$ so that the Lagrangian system we construct will have plenty of integral curves which are not straight lines.

{\bf Case 0b:}\ \ $Z_0^\perp = \sp\{Z_1, Z_2\}$ is not integrable.

Then, $[Z_1,Z_2]$ has a $Z_0$ component and we have to impose $Z_0(V)=0$, so that $V$ must be constant.

It is of some interest to emphasize here that the properties characterizing Case~0 depend on the data only, i.e.\ are independent of the choice of a metric $g$. On the other hand, $g$ is present when it comes to checking whether $\sp\{Z_1, Z_2\}$ is integrable. So we may be able to use the extra freedom of our generalization to avoid the trivial solution for given straight lines.

From now on, we can assume that $\delZ(\phi) \neq 0$ so that the equations (\ref{AZ1}) and (\ref{X}) are well defined. [The case that $\delZ(\phi)=0$ and $\delZ(\psi)\neq 0$ is simply a matter of renaming $\phi$ and $\psi$.] We are then led to look at the bracket of $X$ and $Z_1$, which by the way is equivalent to looking at $\sp\{Z_1, Z_2\}$ again since $X$ belongs to this distribution. The two main cases to be discussed again depend on whether this distribution is integrable or not.

{\bf Case 1:}\ \ $Z_0^\perp = \sp\{X, Z_1\}$ is integrable, say $[X,Z_1] = a\,X + b\,Z_1$ for some functions $a$ and $b$.

It follows from acting with $X$ on (\ref{AZ1}) and $Z_1$ on (\ref{X}) and using the integrability assumption that we must have:
\begin{equation}
\big(X(A) + b\,A\big)(V-\barE) = - A\,X(\barE). \label{case1}
\end{equation}

{\bf Case 1a:}\ \ $X(A) + b\,A \equiv 0$.

So we must have $X(\barE)=0$ in conjunction with the requirement $Z_0(\barE) =0$. Hence, the integrability analysis shifts to $\barE$ and we can distinguish two further subcases.

{\sf Case $1a_1$:}\ \ If the distribution $\sp\{X,Z_0\}$ is integrable, $\barE$ will emerge as an arbitrary function of one variable, which itself will be a specific function of the given $\phi$ and $\psi$ (and also a specific function of variables $(u,v)$ whose exterior derivatives span the co-distribution of $X$).

{\sf Case $1a_2$:}\ \ If  $\sp\{X,Z_0\}$ is not integrable, there will be a third restriction on $\barE$, say $Y(\barE)=0$, where $X$, $Z_0$ and $Y$ are independent vector fields. This implies that $\barE$ is restricted to be an arbitrary numerical constant.

In both of these subcases, however, the integrability requirement (\ref{case1}) for $V$ is satisfied, so we can integrate (\ref{X},\ref{AZ1}), at least in principle, to find a solution for $V$.

{\bf Case 1b:}\ \ $X(A) + b\,A \neq 0$.

We demonstrate in the next section that this case, quite surprisingly, is empty.

{\bf Case 2:}\ \ $\sp\{X, Z_1\}$ is not integrable, say $[X,Z_1] = a\,X + b\,Z_1 + c\,Z_0$.

This time, the integrability computation of (\ref{AZ1}) and (\ref{X}) will give rise to an extra equation, of the form
\begin{equation}
C\,Z_0(V) =  B\,(\barE - V) - X(\barE), \label{case2Z0}
\end{equation}
where we have put
\[
C = cA \qquad\mbox{and}\qquad B = \big(X(A)+b\,A\big)\, A^{-1}.
\]
The extended distribution $\sp\{X,Z_1,Z_0\}$ is integrable by dimension, but the point is that using  $[X, Z_0]$ and $[Z_1, Z_0]$ will produce further necessary conditions, which are linear algebraic expressions in $V$. Explicitly, set
\[
[X, Z_0] = l\, X + m\, Z_1 + n\, Z_0 \quad \mbox{and}\quad	[Z_1, Z_0] = p\, X + q\, Z_1 + r\, Z_0.
\]
Acting with $X$ on (\ref{case2Z0}), using the first of these commutators and replacing first-order derivatives of $V$ by using (\ref{X}), (\ref{AZ1}) and (\ref{case2Z0}), we end up with a relation of the form
\begin{equation}
F_1\, V = G_1. \label{F1V}
\end{equation}
Likewise, using $[Z_1, Z_0]$, we obtain another linear relation, say
\begin{equation}
F_2\, V = G_2. \label{F2V}
\end{equation}
Here, $F_1$ and $F_2$ are `known functions', in the sense that they are expressions involving the functions $A$ and $b$ introduced before, plus derivatives of these functions through the actions of the vector fields $X$, $Z_1$ and $Z_0$, while $G_1$ and $G_2$ depend on $\barE$ and some of its first and second derivatives. Explicitly, we have
\begin{align*}
F_1 &= - (B/C) \big(X(C) + n\, C\big)  +  (C/A)\, m + X(B), \\
G_1 &= F_1 \barE + \big(B + (X(C) + n\, C)/C\big)\, X(\barE)  -  X^2(\barE), \\
F_2 &= q\, C  -  (C/A)\, Z_0(A)  -  B\, \big(r A + (A/C) Z_1(C)\big)  +  A Z_1(B), \\
G_2 &= F_2 \barE  +  \big(r A + 1 + (A/C) Z_1(C)\big) X(\barE)  +  AB\, Z_1(\barE)  -  A Z_1(X(\barE)).
\end{align*}
Again, there is no apparent reason why an algebraic expression for $V$, obtained from (\ref{F1V}) and/or (\ref{F2V}) would automatically satisfy the partial differential equations from which these were derived. One will have to substitute such an expression back into (\ref{X}) and (\ref{AZ1}), leading to higher-order partial differential equations to be satisfied by $\barE$. But to complicate matters, there is even a preliminary compatibility issue between the two expressions (\ref{F1V}) and (\ref{F2V}). In that respect, we can distinguish three subcases, which only slightly differ in the way extra conditions on $\barE$ have to be approached. We should of course not forget that the `known functions' $F_1$ and $F_2$ contain the six parameters of the symmetric matrix $(g_{ij})$. Hence, one of the possible strategies in the search for a solution might be to use these free parameters in order to force the system into a more favourable situation.

{\bf Case 2a:}\ \ $F_1 = F_2 = 0$.

Then $G_1 = 0$  and  $G_2 = 0$ give rise to two second-order conditions on $\barE$. In practice, it remains to be seen whether continuing along these lines offers better chances than trying the `energy free approach'. Also, one could proceed with further integrability considerations concerning the equations $G_1=G_2=0$, but that doesn't really make much sense. Better is to pass to new coordinates, involving $\phi$ and $\psi$ (or in fact any functions of them if more appropriate), plus a suitably selected third variable $\chi$. In fact, as was the case with the `energy free approach' explained in the previous section, it is very likely that by doing this, and knowing that $\barE$ cannot depend on $\chi$, one or both of the equations $G_1=G_2=0$ (being polynomials in $\chi$ for example) immediately split into several conditions, which could then rapidly tell us the form of any nontrivial solution.

Notice that $\barE \!=constant$ is a solution of $G_1=G_2=0$, which should not come as a surprise since the potential $V$ is determined up to a constant. In addition, there is no relation which would indicate that a constant $\barE$ implies a constant $V$. Hence, theoretically, there is a possibility that a non-constant solution of the system of three first-order partial differential equations (\ref{AZ1},\ref{X},\ref{case2Z0}) exists, with $\barE$ in the role of an arbitrary constant.

{\bf Case 2b:}\ \ One of the $F_i$ is zero, say $F_2$.

Then $G_2 = 0$ is a second-order condition on $\barE$ and we can put
\begin{equation}
V = {F_1}^{-1} G_1. \label{Vcase2b}
\end{equation}
As argued before, this $V$ is not guaranteed to solve the original equations (\ref{AZ1},\ref{X}); back substitution in those equations generates more conditions to be satisfied by $\barE$, which will this time even be third-order equations. It again makes no sense to go into details about these equations: the rule of conduct should be to address $G_2=0$ in adapted coordinates $(\phi,\psi,\chi)$, and if a solution can be found, test the $V$ obtained from (\ref{Vcase2b}) by substitution into (\ref{AZ1}) and (\ref{X}). As for the solution  $\barE\!=constant$ in this case, (\ref{Vcase2b}) would imply $V=\barE$ and since we are not in a straight line situation, the conclusion would be that there is no solution.

{\bf Case 2c:}\ \ $F_1\neq 0$ and $F_2 \neq 0$.

This very much looks like a worst case scenario but the examples will show that existence of a solution should not be ruled out. In this case, compatibility between (\ref{F1V}) and (\ref{F2V}) requires that
\begin{equation}
F_2 G_1 = F_1 G_2, \label{F1F2}
\end{equation}
which again is a second-order condition on $\barE$, in which the terms not involving derivatives of $\barE$ cancel out. As in the preceding case, if a constant $\barE$ would turn out to be the only possibility, the conclusion is that there is no solution. Likewise, it is not interesting to look at more conditions on $\barE$ by substitution of (\ref{Vcase2b}) (or equivalently $V = {F_2}^{-1} G_2$) into (\ref{AZ1},\ref{X}). Passing to suitably adapted coordinates $(\phi,\psi,\chi)$ will indeed, in practice, make that (\ref{F1F2}) splits into further equations which are already a headache for Maple\texttrademark \cite{Maple}, or any other computer algebra package one will have to rely on for actually executing the calculations.

To conclude this section, we briefly summarize the motivations behind our theoretical developments. The introduction of a constant symmetric matrix multiplier in the inverse problem, as formulated by (\ref{Vxi1}) is a natural extension of the way the so-called Szebehely problem has mostly been addressed in the literature. It offers six extra parameters which can lead to more general solutions or solutions where there previously were none. Of course, it also leads to extra computational technicalities. Therefore, the practical philosophy we advocate is: (i) if at all possible, try to solve the problem for admissible potentials $V$ via the `energy free approach' explained in Section~2; (ii) alternatively, start the integrability analysis of Section~3, which essentially transfers the problem to equations for the energy function first; (iii) in the course of this process, try to take advantage of the extra free parameters to lead the problem towards a more favourable case of integrability.

We discuss a  number of illustrative examples in sections 5 and 6, including some for which both approaches consistently lead to solutions.  Most of the calculations have been carried out or double checked with Maple\texttrademark.

\section{There is no Case 1b}

The claim we are making here is that whenever the vector fields $X$ and $Z_1$ span an integrable distribution, or equivalently the orthogonal complement of $Z_0$ is integrable, the function $X(A) + b\,A$, as defined in the previous section, is always automatically zero. This sounds like an unrealistic expectation, but we were challenged to pursue this possibility by a result of Shorokhov \cite{Shorokov} in the Euclidean case. Indeed, using a variety of well known identities satisfied by the standard vector calculus operators and (quote) ``omitting tedious calculations'', Shorokhov derives a formula which reads
\begin{equation}
\langle s, \rot p \rangle = \divg \left( \frac{2}{s^2}\,\langle s, \rot s\rangle\,s\right). \label{Shor}
\end{equation}
Here $s$ can be identified with our $Z_0$ and $p$ is proportional to $\delZ$. Integrability of the orthogonal complement of $Z_0$ then corresponds to having $\langle s, \rot s\rangle =0$, and Shorokhov's proof that a potential then always exists is based on the above formula.

In our more general setting, since we want to allow rescaling of $Z_0$ and $X$ for practical, computational purposes, we better check first whether such rescaling could have an effect on the claim we wish to prove. If $\tilde{Z}_0 = \rho\, Z_0$, then $g(\tilde{Z}_0,\tilde{Z}_0) = \rho^2 g(Z_0,Z_0)$ and
\[
\deltZ \tilde{Z}_0 = \rho Z_0(\rho)\,Z_0 + \rho^2 \del0Z Z_0.
\]
It follows that $\deltZ \tilde{Z}_0(\phi) = \rho^2 \del0Z Z_0(\phi)$ and consequently that
\[
\tilde{A} := \onehalf\, \frac{g(\tilde{Z}_0,\tilde{Z}_0)}{\deltZ \tilde{Z}_0(\phi)} = A.
\]
The corresponding definition of $\tilde{X}$ say, reads
\[
\tilde{X} := \deltZ \tilde{Z}_0(\psi)\,Z_1 - \deltZ \tilde{Z}_0(\phi)\,Z_2 = \rho^2 X.
\]
Obviously, integrability of $\sp\{\tilde{X},Z_1\}$ is the same as integrability of $\sp\{X,Z_1\}$ and if $[X,Z_1] = a\,X + b\, Z_1$, then $[\tilde{X},Z_1] = \tilde{a}\,\tilde{X} + \tilde{b}\, Z_1$ with $\tilde{b}=\rho^2 b$. It follows that
\[
\tilde{X}(\tilde{A}) + \tilde{b}\,\tilde{A} = \rho^2\big(X(A) + b\,A\big),
\]
and this is also the conclusion if we independently rescale $X$, with or without a rescaling of $Z_0$. Hence, rescaling causes no threat for the property we want to obtain.

Let us now first try to reformulate the problem under consideration in more geometrical terms. To this end, if we put $Y:= A\,Z_1$, integrability of $Z_0^\perp$ equally means that $[X,Y] = \bar{a}\,X + \bar{b}\,Y$ for some functions $\bar{a}, \bar{b}$. With the starting equations (\ref{X}) and (\ref{AZ1}) now in the form $X(V)=0,\ Y(V)=V-\barE$, the commutator requirement (\ref{case1}) takes the form $\bar{b}\,V = \bar{b}\,E - X(E)$. So in this representation, we want to prove that $\bar{b}=0$. But if a vector field has only an $X$-component, it is orthogonal to both $Z_0$ and $\delZ$. Hence, our goal is to show that
\begin{equation}
g([X,Y], Z_0) = 0 \qquad \Longrightarrow \qquad g([X,Y], \delZ) =0, \label{goal}
\end{equation}
and it will be convenient to approach this via the dual version of Frobenius' theorem. Putting
\begin{equation}
Z_0 \hook g = \alpha_0, \label{alpha0}
\end{equation}
we know from (\ref{Zprops}) that $\alpha_0$ annihilates $Z_0^\perp$, so that
\begin{equation}
g([X,Y], Z_0) = 0 \qquad \Longleftrightarrow \qquad d\alpha_0\wedge\alpha_0 =0. \label{dualFrob}
\end{equation}

\begin{lem} \ \ $\del0Z\alpha_0 = i_{Z_0}d\alpha_0 + \onehalf d\,\big(g(Z_0,Z_0)\big)$.
\end{lem}
{\sc Proof:}\ We have, using the fact that the $g_{kl}$ are constant,
\begin{align*}
\big(\del0Z\alpha_0\big)_k &= g_{kl}Z_0^j\fpd{Z_0^l}{x^j} \\
&= Z_0^j\left(g_{kl}\fpd{Z_0^l}{x^j} - g_{jl}\fpd{Z_0^l}{x^k}\right) + \frac{1}{2}\fpd{}{x^k}(g_{jl}Z_0^j Z_0^l),
\end{align*}
from which the result follows. \qed

Now put
\begin{equation}
\beta_0 := \frac{2}{g(Z_0,Z_0)}\del0Z\alpha_0. \label{beta0}
\end{equation}

\begin{lem} {} $\displaystyle d\beta_0\wedge\alpha_0 = - \frac{2}{g(Z_0,Z_0)^2}\,d\,\big(g(Z_0,Z_0)\big) \wedge i_{Z_0}(d\alpha_0\wedge\alpha_0) + \frac{2}{g(Z_0,Z_0)}\,\lie{Z_0}(d\alpha_0\wedge\alpha_0)$.
\end{lem}
{\sc Proof:}\ Using the result of Lemma~1, we easily find that
\[
d\beta_0 = - \frac{2}{g(Z_0,Z_0)^2}\,d\,\big(g(Z_0,Z_0)\big) \wedge i_{Z_0}d\alpha_0 + \frac{2}{g(Z_0,Z_0)}\,\lie{Z_0}d\alpha_0.
\]
Wedging this with $\alpha_0$, the difference between what we get and what we want to prove is a sum of three terms, two of which cancel each other, while the remaining one is proportional to $d\alpha_0\wedge i_{Z_0}d\alpha_0$. But this last term is zero also, because it is half of the contraction with $Z_0$ of $d\alpha_0\wedge d\alpha_0$, which is a $4$-form in dimension 3. \qed

Remark: Computing $d\beta_0\wedge\alpha_0$ can be seen to be exactly the analogue of what is the left-hand side in the result (\ref{Shor}) obtained by Shorokhov.

\begin{thm} Whenever a non-singular symmetric multiplier matrix $g_{ij}$ can be found, such that $Z_0^\perp = \sp\{Z_1,Z_2\} = \sp\{X,Y\}$ is an integrable distribution, the generalized Szebehely problem expressed by the conditions (\ref{X}, \ref{AZ1}) has a solution for the potential, whereby the corresponding energy function $\barE$ will be either an arbitrary function of one specific variable, or a numerical constant.
\end{thm}
{\sc Proof:}\ It directly follows from Lemma~2 that the integrability assumption $d\alpha_0\wedge \alpha_0=0$ implies that $d\beta_0\wedge\alpha_0=0$. With, for example, $\{X, Y, Z_0\}$ as a local basis of vector fields, knowing that $\alpha_0(X) = \alpha_0(Y)=0$, it immediately follows that
\[
d\beta_0\wedge\alpha_0=0 \qquad \Longleftrightarrow \qquad d\beta_0(X,Y)=0.
\]
In addition, $g(X,\delZ)=0$ says that $\beta_0(X)=0$, while
\[
\beta_0(Y) = \frac{2}{g(Z_0,Z_0)}\,\del0Z\alpha_0 \left(\onehalf \frac{g(Z_0,Z_0)}{\delZ(\phi)}\,Z_1\right) = \frac{1}{\delZ(\phi)}\, g(Z_1,\delZ) \equiv 1.
\]
Hence $d\beta_0(X,Y) = - \beta_0([X,Y])$, which is proportional to $g([X,Y],\delZ)$. The overall conclusion, taking also (\ref{dualFrob}) into account, now is that the implication (\ref{goal}) is true. Therefore, Case~1b is void, so that whenever the distribution $Z_0^\perp$ is integrable, we are in Case~1a. The result then follows from our analysis of Case~1a in the preceding section. \qed

\section{A broader look at some examples from the literature}

Shorokhov, in his integrability analysis \cite{Shorokov}, discusses four examples to illustrate some of the cases in his classification. Since that classification is quite different from the way we have approached the integrability problem, it is appropriate that we check where these examples fit into our scheme. In fact, two of Shorokhov's examples (taken over from previous publications) concern straight lines, so we single them out first. In what follows we won't be explicit about the domains of the various functions involved.

{\sc Example 1:} \ \ $\phi = x/z$,\ \ $\psi=y/z$.

After rescaling, we have that
\[
Z_0 = x\fpd{}{x} + y\fpd{}{y} + z\fpd{}{z}, \quad \mbox{and} \quad \delZ = Z_0.
\]
It is easy to verify that, for the standard choice $g_{ij}=\delta_{ij}$, we are in Case~0a and the potential, as reported in \cite{Shorokov}, can be any function of $x^2+y^2+z^2$, whereas there is no restriction on the energy function $\barE$ as a function of $\phi$ and $\psi$. For comparison now, we look at what happens when we allow a more general, non-singular $g$. Note for a start that, since the problem is reduced to solving the equations $Z_1(V)=Z_2(V)=0$, we can somewhat simplify the calculations again by rescaling $Z_1$ and $Z_2$ as well. It turns out that $\sp\{Z_1,Z_2\}$ is integrable for any choice of $g$, i.e.\ that we remain in Case~0a no matter what constant, kinetic energy metric we select. Correspondingly, for an arbitrary $g$, the potential function $V$ can be taken as an arbitrary function of the quadratic expression $g_{ij}x^ix^j$. Needless to say, the given 2-parameter family of straight lines constitutes only a particular subset of the integral curves of the resulting Lagrangian system.

{\sc Example 2:}\ \ $\phi=x/y$, \ \ $\psi= y+z$.

Upon appropriate rescaling,
\[
Z_0 = - x\fpd{}{x} - y\fpd{}{y} + y\fpd{}{z}, \quad \mbox{and} \quad \delZ = - Z_0.
\]
This time, with  $g_{ij}=\delta_{ij}$, we are in Case~0b. Hence, the only solution is the trivial one $V\!= constant$ and we are looking at the free particle system whose integral curves all are straight lines. However, this is a first example of a case where our broader picture can bring us into a more favourable situation of integrability, with a less trivial solution for admissible potentials. Indeed, if we take $g_{13}=0$ and $g_{23}=g_{33}$, it turns out that we move up to Case~0a. To be a bit more explicit,  appropriately rescaled versions of $Z_1$ and $Z_2$ then read
\[
Z_1 = \big(g_{12}x + (g_{22}-g_{33})y\big)\fpd{}{x} - (g_{11}x + g_{12}y)\Big(\fpd{}{y} - \fpd{}{z}\Big), \qquad Z_2 = \fpd{}{z},
\]
and it is clear that these vector fields commute. The solution now reads that $V$ can be an arbitrary function of the expression $g_{11}x^2 + (g_{22}-g_{33})y^2 + 2g_{12}xy$.

{\sc Example 3:}\ \ $\phi=xz$, \ \ $\psi=yz$.

It is straightforward to verify that for the Euclidean metric $g_{ij}=\delta_{ij}$, this is an example belonging to the most favourable Case~$1a_1$, so we cannot expect to do much better by allowing a more general $g$. Yet, it is of interest to see whether more general solutions exist and we will take the opportunity to illustrate some of the technical problems to which we referred before as a motivation for our integrability analysis.

For a start, after the usual rescaling, the vector fields $Z_0$ and $\delZ$ read:
\[
Z_0 = -x\fpd{}{x} - y\fpd{}{y} + z\fpd{}{z}, \qquad \delZ = x\fpd{}{x} + y\fpd{}{y} + z\fpd{}{z},
\]
whereas
\begin{align*}
X &=  \big((g_{12}g_{33}-g_{13}g_{23})\,x + (g_{22}g_{33} - g_{23}^2)\,y\big) \fpd{}{x} \\
& \hspace*{1cm} + \big( (g_{13}^2 - g_{11}g_{33})\,x + (g_{13}g_{23} - g_{12}g_{33})\,y\big) \fpd{}{y} \\
& \hspace*{2cm} + \big( (g_{11}g_{23} - g_{12}g_{13})\,x + (g_{12}g_{23} - g_{13}g_{22})\,y \big) \fpd{}{z}.
\end{align*}
We can actually solve the equation $X(V)=0$ in all generality here. The implication is that $V$ should be a function of, say
\begin{align*}
u &= g_{13}x + g_{23}y + g_{33}z, \\
v &= \onehalf (g_{11}g_{33} - g_{13}^2)\,x^2 + (g_{12}g_{33} - g_{13}g_{23})\,xy + \onehalf (g_{22}g_{33} - g_{23}^2)\,y^2.
\end{align*}
We did not succeed, however, in continuing along the `energy free approach' by solving the second-order condition (\ref{Z0Z1}). Hence, we are led into the integrability scheme and it is easy to check that we can force the system into Case~1, by taking $g_{13}=g_{23}=0$. As by now expected, this simplification also results in $X(A) + bA=0$, so that we are in Case~1a. Moreover, with the reduced expression of $X$, we now have that $[X,Z_0]=0$, meaning that, even better, we are in Case~$1a_1$. The conditions on $\barE$ require that $Z_0(\barE)=X(\barE)=0$ and tell us that $\barE$ can be an arbitrary function of one variable, which itself can be expressed as function of $\phi$ and $\psi$. We choose to represent the solution for $\barE$ in the following form:
\[
\barE = F\Big(\onehalf\,g_{33}\,(g_{11}\phi^2 + 2g_{12}\phi\psi + g_{22}\psi^2)\Big).
\]
Having done that, we are guaranteed that the equations (\ref{AZ1}) and (\ref{X}) have a solution for $V$. Of course, such a $V$ will still have to be a function of the (reduced) variables $(u,v)$ mentioned before. It is therefore of interest that the representation we choose for $\barE$ can easily be expressed as function of $(u,v)$ as well. Even more important is that we try to optimize the choice of $(u,v)$ variables now. For example, if we manage to pass to an equivalent set $(u_1,v_1)$ such that $Z_1(v_1)=0$, then the remaining equation (\ref{AZ1}) will be greatly simplified. Such an equivalent set exists here and is given by
\begin{align*}
u_1 &= \onehalf (g_{11}x^2 + 2 g_{12}xy + g_{22}y^2 + g_{33}z^2), \\
v_1 &= \onehalf (g_{11}x^2 + 2 g_{12}xy + g_{22}y^2 - g_{33}z^2).
\end{align*}
The expression for $\barE$ above then can be written as
\[
\barE = F\Big(\onehalf(u_1 + v_1)(u_1 - v_1)\Big),
\]
and equation (\ref{AZ1}) reduces to
\[
u_1\fpd{V}{u_1} = V - F\Big(\onehalf(u_1 + v_1)(u_1 - v_1)\Big).
\]
The solution is given by
\[
V = u_1\,\left[ G(v_1) - \int \frac{1}{u_1^2}\, F\big(\onehalf(u_1^2 - v_1^2)\big)\, du_1 \right]
\]
with $F$ and $G$ arbitrary functions (not both zero) of the indicated arguments. For the particular case that $g_{12}=0$ and $g_{11}=g_{22}=g_{33}=1$, we recover here exactly the solution reported in \cite{Shorokov}.

{\sc Example 4:}\ \ $\phi=xyz + x + y$, \ \ $\psi=z$.

We have
\[
Z_0 = (xz+1)\fpd{}{x} - (yz+1)\fpd{}{y},
\]
and a rescaled version of $X$ reads
\[
X = (g_{13}g_{22} - g_{12}g_{23})\fpd{}{x} + (g_{11}g_{23}-g_{13}g_{12})\fpd{}{y} -(g_{11}g_{22}-g_{12}^2)\fpd{}{z}.
\]
One can verify that it is impossible to select a non-singular $(g_{ij})$  which would make the orthogonal complement of $Z_0$ integrable, hence we are condemned to face the difficulties of Case~2. We note in passing that the equation $X(V)=0$ can be integrated here as well and identifies variables $(u,v)$ as follows:
\begin{align*}
u &= (g_{11}g_{23} - g_{12}g_{13})\,x + (g_{12}g_{23} - g_{13}g_{22})\,y, \\
v &= (g_{11}g_{22} - g_{12}^2)\,x + (g_{13}g_{22} - g_{12}g_{23})\,z.
\end{align*}
Unfortunately, continuing along the path of the `energy free approach' looks impossible without further restrictions on $g$. What we will observe with this example, however, is that the Case~2 integrability analysis may actually suggest or impose suitable restrictions.

Following the Case~2 scheme now, there is no apparent incentive to enforcing that either $F_1$ or $F_2$ be zero, which means that we shall start by addressing a Case~2c situation. The first obstruction for a solution then is (\ref{F1F2}), a second-order condition on $\barE$ which requires passing to adapted coordinates $(\phi,\psi,\chi)$. We make the choice $\chi=y$, by which the inverse transformation becomes
\[
x= \frac{\phi - \chi}{\psi\chi +1}, \quad y = \chi, \quad z=\psi.
\]
Expressing the condition (\ref{F1F2}) in the new variables must be handled with great care. In doing so we find that the resulting condition is a polynomial in $\chi$ of degree 29. In other words, we are looking here, not at a single second-order condition on a function $\barE(\phi,\psi)$, but at a set of 30 such conditions which must be imposed step by step. We briefly sketch the technicalities which follow. From the coefficient of the highest-order term, it follows that if $g_{22}\neq 0$, $\barE$ will have to be of the form
\[
\barE(\phi,\psi) = f_1(\psi) + \frac{f_2(\psi)}{\phi\psi +1}.
\]
The terms of order 28, however, immediately restrict $f_1$ to be constant and thus redundant. This has the effect that the next non-vanishing condition comes from the terms of order 25, which splits into two conditions again because it is a polynomial of degree 1 in $\psi$ now. The solution of these two conditions reads
\[
f_2(\psi) = C_1 \psi^2 + C_2\psi\big((g_{12}g_{23} - g_{13}g_{22})\psi^2 + g_{11}g_{22} - g_{12}^2\big).
\]
Now things become even more cumbersome because each of the subsequent conditions will split into many more since they each become polynomials in $\phi$ and $\psi$. Looking at a suitable subset of these conditions (still under the assumption that $g_{22}\neq 0$), we find that only the following subcases are worth a further analysis: either $C_2=0$ or $g_{11}=g_{12}=0$. The case $g_{11}=g_{12}=0$ turns out to make $F_1=0$ so that we are actually in Case~2b. It can further be verified that also $G_1=0$, so there are no further restrictions on the expression for $\barE$ obtained so far. It follows that $V$ should be $G_2/F_2$. Such a $V$ actually satisfies $X(V)=0$, but unfortunately not the other initial requirement (\ref{AZ1}), hence there is no solution. It is interesting to observe here in passing that with $g_{11}=g_{12}=0$, the vector field $X$ becomes proportional to $\partial/\partial x$ which implies $V=V(y,z)$. We can then simply pursue the `energy free approach' and it leads to the conclusion that $V$ must be constant, confirming this way that there is no solution. With the alternative choice $C_2=0$, going back to the requirement (\ref{F1F2}) now and cutting short a rather long story, an investigation of all conditions coming from terms of degree 24 and 23 soon leads to the conclusion that (maintaining the assumption $g_{22}\neq 0$), we will have to put $g_{11}=g_{22}$ and $g_{13}=g_{23}$. Meanwhile, with $C_2=0$, the expression for $\barE$ we obtained so far is reduced to
\[
\barE = \frac{C_1 \psi^2}{\phi\psi +1}.
\]
Fortunately, the massive number of remaining terms in the compatibility condition (\ref{F1F2}) all vanish now (while by far this time, the functions $F_1$ and $F_2$ are not zero). Knowing that (\ref{F1V}) and (\ref{F2V}) are now compatible, it remains to use any of these relations to define $V$ and check whether such a $V$ happens to satisfy the equations (\ref{X}) and (\ref{AZ1}). This turns out to be the case without further restrictions on the multiplier $g$. Expressed in the original variables, the solution for the potential is of the form
\[
V = \frac{C}{(x-y)^2},
\]
whereby the relation between the constants $C$ and $C_1$ is given by $C= 2C_1 (g_{12}-g_{22})/g_{22}$. It is worth observing that with $g_{11}=g_{22}$ and $g_{13}=g_{23}$, a choice which was dictated by the analysis of equation (\ref{F1F2}), we could again forget the integrability analysis and go back to the `energy free approach'. The $(u,v)$ variables mentioned above are no longer valid though. The vector field $X$ can be rescaled to
\[
X = g_{23}\Big(\fpd{}{x} + \fpd{}{y}\Big) - (g_{12} + g_{22})\fpd{}{z},
\]
which shows that appropriate variables for $V$ now are: $u= x-y$ and $v= (g_{12} + g_{22})x + g_{23}z$. The second-order equation (\ref{Z0Z1}) then implies that $V$ must be the function of $u$ specified above.

For completeness, it remains to investigate the case that $g_{22}=0$, since that was excluded at the very first stage of our integrability analysis. This reduces the order of the polynomial in $\chi$ from 29 to 19, but the remaining conditions are still so strong that the final conclusion is: no solution!

A final comment is in order about this example. The solution we have obtained for $V$ is exactly the same as the one found by Shorokhov \cite{Shorokov} even though the restrictions on our $g$ are not forcing it to be proportional to the standard Euclidean metric. So, in some sense, our generalization fails to produce more general admissible potentials here. Instead, it has the curious effect that with the same potential, a broader class of kinetic energy terms is allowed to arrive at a Lagrangian system which has the given family of curves sitting in its set of integral curves. Explicitly, the more general kinetic energy is of the form
\[
T = \onehalf g_{22}(\dot{x}^2 + \dot{y}^2) + \onehalf g_{33}\dot{z}^2 + g_{12}\dot{x}\dot{y} + g_{23}(\dot{x} +\dot{y})\dot{z}.
\]

{\sc Example 5:}\ \ $\phi=\onehalf (x^2 + y^2)$, \ \ $\psi=z/x$.

This example has been discussed (for the Euclidean case) in a number of papers. It is the first example in \cite{VE83}, but as mentioned before, these authors start from a preassigned energy function for which no justification is given. A better treatment is found in \cite{Anisiu2005}, where the general solution is obtained via the `energy free approach'. Let us first sketch how this presents itself with the notations we have introduced. After an appropriate rescaling, we have
\[
Z_0 = xy \fpd{}{x} - x^2 \fpd{}{y} + yz \fpd{}{z}, \quad\mbox{and}\quad X= -z\fpd{}{x} + x \fpd{}{z},
\]
whereby $X$ in fact is proportional to $Z_2$ here. Solving the equation $X(V)=0$ identifies $V$ as a function of say $u=x^2 +z^2$ and $v=y$. Putting in addition $w=\psi=z/x$ completes a set of new variables which turns out to be viable for expressing the further objects of interest into such coordinates.  Imposing the second-order condition (\ref{Z0Z1}) subsequently pins down the potential, which (up to an additive constant) is given by
\begin{equation}
V(u,v) = C_1 (u + v^2) + C_2/v^2 - C_3/u, \qquad C_i \mbox{ constant}. \label{A2V}
\end{equation}
The corresponding energy function $\barE$, which as we know must also be expressible as a function of $\phi$ and $\psi$ is found to be
\begin{equation}
\barE = 2\phi(\psi^2 +2)\,C_1 - \frac{\psi^2}{2\phi}\,C_2 - \frac{\psi^2}{2\phi(\psi^2 +1)^2}\,C_3. \label{A2E}
\end{equation}
It is interesting that this solution can also be obtained by following the integrability scheme. We find ourselves in Case~2c then so that the first problem to address is solving the second-order equation (\ref{F1F2}) for $\barE$. For that, it is appropriate to use coordinates $(\phi,\psi,\chi)$ adapted to $\barE$, for which one can simply take $\chi=y$. This way, the equation (\ref{F1F2}) becomes a polynomial in $\chi$, with only even power terms up to order 6. Hence, there is a set of four second-order equations to be solved, but this can be done and the solution is found to be compatible with (\ref{A2E}). However, in principle, this is not the end of the story. Our scheme tells us that we have to put $V={F_1}^{-1}G_1 = {F_2}^{-1}G_2$ and test whether this $V$ satisfies the equations (\ref{X}) and (\ref{AZ1}). This may lead in general to many further restrictions, but here it doesn't and we have indeed obtained the solution for $V$.

Let us now investigate whether more general potentials can be obtained by introducing our constant multiplier $g_{ij}$. That has no effect on the $Z_0$ mentioned above, but $X$ becomes
\begin{align*}
X &= \Big((g_{12} g_{23} - g_{13} g_{22})\, x + (g_{23}^2 - g_{22} g_{33})\, z\Big) \fpd{}{x} \\
	& \hspace{1cm}	\mbox{} + \Big((g_{12} g_{13} - g_{11} g_{23})\, x + (g_{12} g_{33} - g_{13} g_{23})\, z\Big) \fpd{}{y} \\
	& \hspace{2cm}	\mbox{}	+ \Big((g_{11} g_{22} - g_{12}^2)\, x + (g_{13} g_{22} - g_{12} g_{23}) \, z\Big) \fpd{}{z}.
\end{align*}
It is fairly easy to verify that there is no non-singular $g$ which produces a Case~1 situation. So either we face the Case~2 complications or we try the energy free approach. For a start, observe that we can solve the equation $X(V)=0$ without restrictions on $g$. Corresponding $(u,v)$-variables are
\begin{align}
u &= \big(g_{11}g_{22} - g_{12}^2\big)x^2 +2\big(g_{13}g_{22} - g_{12}g_{23}\big) xz + \big(g_{22}g_{33} - g_{23}^2\big) z^2, \label{uA2} \\
v &= g_{12}x + g_{22}y + g_{23}z. \label{vA2}
\end{align}
Unfortunately, it looks impossible to select a third variable $w$ in such a way that we could even set up the second-order condition (\ref{Z0Z1}). The alternative approach via $\barE$ has the advantage that adapted coordinates $(\phi,\psi,\chi)$ are not affected by the presence of $g$. Yet again, we were unable, even with the assistance of Maple\texttrademark, to get anything out of the second-order condition (\ref{F1F2}) for general $g$. Even the expressions for the known functions $F_1$ and $F_2$, for example, run over several pages. So, we have to settle for some particular cases.

Inspired by the general solution (\ref{A2V}) for the Euclidean case, let us try under what conditions on $g$ a $V$ of the form $V=C_3/u$ solves the second-order requirement (\ref{Z0Z1}). The conclusion is that this is the case provided we restrict $g$ by $g_{12}=g_{23}=0$. Exactly the same restriction makes that also a $V$ of the form $C_2/v^2$ works, which is no surprise. On the other hand, it turns out that $V=C_1(u+v^2)$ is an admissible potential without any restriction on $g$. Turning the arguments around, if we set $g_{12}=g_{23}=0$ from the outset, we can find that the general solution of (\ref{Z0Z1}) is a linear combination of the above three particular solutions. Remember that the function $u$ now is more general than in the standard Euclidean case and this is made possible by choosing a kinetic energy of the form
\[
T= \onehalf (g_{11}\dot{x}^2 + g_{22}\dot{y}^2 + g_{33}\dot{z}^2) + g_{13}\dot{x}\dot{z}.
\]
Finally, let us attempt to find a particular solution which can manifestly not be obtained with a standard kinetic energy. To that end, we put $g_{11}=g_{22}=g_{33}=0$. Then, the homogeneous partial differential equation $X(V)=0$ identifies variables
\[
u = g_{12}x + g_{23}z, \quad \mbox{and} \quad v= g_{12} xy + g_{13}xz + g_{23}z.
\]
If we put $w = \psi = z/x$ to complete a local coordinate change, the second-order condition for $V$ turns out to generate a polynomial in $w$ of degree 13. Hence, no less than 14 second-order conditions have to be satisfied by a function $V(u,v)$. But there is a solution, namely $V= C_1 v$ (always up to an additive constant). The corresponding energy function, expressed in its own natural variables, reads $\barE = - 2g_{13}C_1 \phi\psi$.

\section{Some new examples}

As far as we know, the following example has not been mentioned in the extensive literature before:
\[
\phi = (x+y)z, \qquad \psi = xy.
\]
The vector fields $Z_0$ and $\delZ$ are given by
\begin{align*}
Z_0 &= - x(x + y) \fpd{}{x} + y(x + y) \fpd{}{y} + z(x - y) \fpd{}{z}, \\
\delZ &= 2 x^2(x + y) \fpd{}{x} + 2 y^2(x + y) \fpd{}{y} - 4 xyz \fpd{}{z} .
\end{align*}
With an arbitrary $g$, the other vector fields of interest are too long to list up here and there is no chance to integrate the fundamental equation $X(V)=0$. With, for example, $\chi=y$ as additional variable, we have an easily invertible coordinate change, so the approach via the energy condition $F_1 G_2 = F_2 G_1$ seems to offer better chances for success. But it takes a lot of effort to even control the expression swell in Maple\texttrademark and the result is bound to be a polynomial in $\chi$ of very high degree. In short, we were unable to get results without a priori simplifying assumptions on $g$. Let us then look at two complementary situations.

First assume that $g$ is diagonal, i.e.\ $g_{12}=g_{13}=g_{23}=0$. The condition referred to above then `reduces' (after eliminating denominators) to a polynomial of degree 76 in $\chi$. The coefficients of the highest and lowest degree terms produce the same equation with $g_{11}$ and $g_{22}$ interchanged. Therefore, we must have $g_{11}=g_{22}$ and since there is an overall scaling freedom, we can actually set them equal to 1 without loss of generality. Bringing the coefficient of the next to highest order terms into the picture, we learn that $\barE$ is bound to be a function of the form $a\phi^2+b\psi+c$, with $a,b,c$ as yet arbitrary constants. But the remaining terms in the condition $F_1 G_2 = F_2 G_1$ soon impose that $a$ and $b$ should be zero and hence that $\barE$ must be constant, which in a Case~2c situation means that there is no solution. We conclude that the problem has no solution for a diagonal $g$, which of course contains the standard Euclidean case.

Secondly, as a kind of complementary hypothesis, assume that $g_{11}=g_{22}=g_{33}=0$. Attempts to solve the equation $X(V)=0$ are still unsuccessful. The energy approach this time leads to a polynomial expression of degree 56, and the highest and lowest order coefficients turn out to vanish identically provided we take $g_{13}=g_{23}$, which as before can then be set equal to 1 without loss of generality. However, at this point we appear to be in a Case~1 setting, so that we better take a step backwards and look at this situation from a different angle.

Without any a priori assumptions on the metric $g$, let us try to implement the strategy of forcing the system into a Case~1 situation. For that purpose, we compute the bracket $[X,Z_1]$ in the local frame $\sp\{X,Z_1,Z_0\}$ and require that the coefficient of $Z_0$ be zero. It turns out that this will be the case, provided that
\begin{align*}
g_{23} (g_{11}+g_{12}) &= g_{13} (g_{22}+g_{12}), \\
g_{33} (g_{12} - g_{11}) &= g_{13} (g_{23} - g_{13}), \\
g_{33} (g_{22} - g_{12}) &= g_{23} (g_{23} - g_{13}) .
\end{align*}
One can show that there is a non-singular $g$ satisfying these conditions if and only if
\[
g_{13} = g_{23} ,\quad	g_{33} = 0 ,\quad g_{22} = g_{11} \qquad	\mbox{with}\  g_{11} \neq g_{12}.
\]
Again, using the scale freedom in $g$, we set $g_{13}=g_{23}=1$ and thus cover the situation encountered in the preceding paragraph. With these simplifications, we now list the expressions for the various objects that matter for the subsequent considerations. The vector fields $X$ (after rescaling) and $Z_1,Z_2$ are given by
\begin{align*}
X &= - \Big((g_{11} - g_{12})xy(x+y)^3 + z(x - y)^3(x+y)\Big) \fpd{}{x} \\
& \hspace{1cm}	\mbox{} - \Big((g_{11} - g_{12})xy(x+y)^3 - z(x - y)^3 (x+y)\Big) \fpd{}{y} \\
& \hspace{2cm}	\mbox{} + \Big((g_{11}^2 - g_{12}^2)xy(x+y)^3 + (g_{11} - g_{12})z(x+y)^2(x^2+y^2 - 4xy)\Big) \fpd{}{z}, \\
Z_1 &= \onehalf (x+y) \bigg( \fpd{}{x} + \fpd{}{y}\bigg) + \Big(z - \onehalf (g_{11} + g_{12})(x+y)\Big) \fpd{}{z}, \\
Z_2 &= \onehalf \frac{x - y}{g_{12} - g_{11}} \bigg( \fpd{}{x} - \fpd{}{y}\bigg) +\onehalf (x+y) \fpd{}{z} .
\end{align*}
In addition, we have for the function $A$ defined by (\ref{A}) and the coefficient $b$ of $Z_1$ in the bracket $[X,Z_1]$ the following expressions:
\begin{align*}
A &= - \onehalf + \onequarter [g_{11}(x+y)^3 - 2(g_{11}+g_{12}) xy(x+y)]/z(x - y)^2, \\
b &= (xy/z)(g_{11}^2 - g_{12}^2) (x+y)^3 + (g_{11} - g_{12})(x+y)^4.
\end{align*}
One can verify that $X(A) + b\,A =0$, as it should, so that we are in Case~1a indeed. Following our general scheme, we next investigate integrability of $\sp\{X,Z_0\}$. It turns out that this distribution is not integrable because $g_{11}\neq g_{12}$. So we are in Case~1$a_2$ and conclude that $\barE$ must be constant. It remains to integrate the system of equations (\ref{X}) and (\ref{AZ1}), and Maple\texttrademark helps us to conclude that the solution $V$ can be written in the form $V = u F(v) + \barE$,
where $F$ is an arbitrary function and the variables $u$ and $v$ are (in the domain of their definition) given by
\begin{align*}
u &= \frac{g_{11}}{(g_{11}+g_{12})(x-y)^2} - \frac{2xy}{(x^2 - y^2)^2} - \frac{2z}{(g_{11} + g_{12})(x+y)^3},  \\
v &= \frac{z}{x+y} + \onehalf (g_{11}+g_{12})\,\ln (x+y)  + \onehalf (g_{11}-g_{12})\,\ln (x-y).
\end{align*}
So this is definitely a non-trivial example where a potential does not exist if one sticks to the traditional picture with a standard Euclidean metric, whereas we have obtained a $V$, provided we take the kinetic energy to be
\[
T = \onehalf g_{11} (\dot{x}^2 + \dot{y}^2) + g_{12} \dot{x}\dot{y} + \dot{x}\dot{z} + \dot{y}\dot{z}.
\]

For some more new examples now, we will avoid the very technical complications of Case~2 situations and address directly the approach of using the extra freedom of a $g$ to end up in a Case~1 situation. This part of the computations at least is quite straightforward and when successful, we know from Section~4 that a potential can be obtained. The idea now is to keep some arbitrariness in the a priori selection of the functions $\phi$ and $\psi$ and hope that imposing integrability of $\sp\{X,Z_1\}$ will be possible without ending up with the conclusion that $g$ will be singular. Let us take
\[
\phi = \onehalf xy^2, \qquad \psi = z\,f(y),
\]
with $f$ as yet undetermined. The condition to be in Case~1 is a polynomial expression in $x$ and $z$
which is of degree 1 in both variables. Hence, it splits into 4 conditions, which read:
\begin{alignat*}{2}
& g_{12}g_{23} - g_{13}g_{22} = 0 &\qquad\mbox{or}\qquad & y\,f' - 2f =0, \\
& g_{11}g_{23} - g_{12}g_{13} = 0 &\qquad\mbox{or}\qquad & y\,f' + f =0, \\
& g_{12}g_{33} - g_{13}g_{23} = 0 &\qquad\mbox{or}\qquad & y\,f f'' - y\,{f'}^2 - 2f f'=0, \\
& g_{13}^2 - g_{11}g_{33} = 0 &\qquad\mbox{or}\qquad & y\,f f'' - y\,{f'}^2 + f f'=0.
\end{alignat*}
A careful analysis of these conditions, keeping in mind of course that $g$ must be non-singular, ultimately gives rise to the following four distinct possibilities (where $f$ always is determined up to an irrelevant multiplicative constant).
\begin{enumerate}
\item $g_{12}=g_{13}=0$ ($g_{23}\neq 0$), with $f= 1/y$.
\item $g_{12}=g_{13}=g_{23}=0$, with $f= y^k$ ($k$ arbitrary).
\item $g_{11}=g_{13}=g_{23}=0$, with $f= e^{ky^3}$ ($k$ arbitrary).
\item $g_{13}=g_{23}=0$ ($g_{11}\neq 0$, $g_{12}\neq 0$), with $f= 1$.
\end{enumerate}
It is instructive to verify again that in each of these cases, we have $X(A) + b\,A=0$ indeed, so that we are in Case~1a. The computation of the solution for the potential in each of the above situations is fairly straightforward.

\subsubsection*{Acknowledgements}  W.S.\ acknowledges support from the Czech Science Foundation under grant GACR 14-02476S ``Variations, Geometry and Physics". G.P.\ thanks the Department of Mathematics at Ghent University for its hospitality.

\end{document}